\documentclass[
    ,final            
  ]
  {aipproc}

\layoutstyle{6x9}
\usepackage{psfrag}
\usepackage{wasysym}

\newcommand{\beq}{\begin{equation}}
\newcommand{\eeq}{\end{equation}}
\newcommand{\bea}{\begin{eqnarray}}
\newcommand{\eea}{\end{eqnarray}}

\def\tit#1#2#3#4#5{{#1}{\bf #2}, #3 (#4)}
\def\jmp{J.\ Math.\ Phys.\ }

\def\prl{Phys.\ Rev.\ Lett.\ }

\def\prb{Phys.\ Rev.\ B\ }

\def\sci{Science\ }

\begin{document}

\title{From exotic phases to microscopic Hamiltonians}


\author{R. Moessner}{
  address={Laboratoire de Physique Th\'eorique de l'Ecole Normale Sup\'erieure;
CNRS-UMR 8549;
24, rue Lhomond; 75231 Paris Cedex 05; France}
}

\author{K. S. Raman}{
 address={Department of Physics, University of Illinois at
Urbana-Champaign, Urbana, IL 61801, USA} 
}

\author{S. L. Sondhi}{
 address={Department of Physics, Princeton University,
Princeton, NJ 08544, USA} 
}

\begin{abstract}
We report recent analytical progress in the quest for spin models
realising exotic phases. We focus on the question of
`reverse-engineering' a local, SU(2) invariant S=1/2 Hamiltonian to
exhibit phases predicted on the basis of effective models, such as
large-$N$ or quantum dimer models. This aim is to provide a
point-of-principle demonstration of the possibility of constructing
such microscopic lattice Hamiltonians, as well as to complement and
guide numerical (and experimental) approaches to the same question. In
particular, we demonstrate how to utilise peturbed Klein Hamiltonians
to generate effective quantum dimer models. These models use local
multi-spin interactions and, to obtain a controlled theory, a
decoration procedure involving the insertion of Majumdar-Ghosh
chainlets on the bonds of the lattice. The phases we thus realise
include deconfined resonating valence bond liquids, a devil's
staircase of interleaved phases which exhibits Cantor deconfinement,
as well as a three-dimensional $U(1)$ liquid phase exhibiting photonic
excitations. 

\end{abstract}

\maketitle


\section{Introduction}
In the early 1970s, Anderson and Fazekas \cite{andersonfazekas}
proposed that the $S=1/2$ quantum Heisenberg antiferromagnet on the
triangular lattice should exhibit a new type of phase, the resonating
valence bond (RVB) liquid. Unlike a conventional N\'eel phase, the RVB
liquid would retain the full symmetry of the spin Hamiltonian, and
thus neither break spatial nor time-reversal symmetries. In addition,
an odd number of sites per unit cell in such a state also implies the
existence of an unambiguous Mott insulator not adiabatically connected
to a simple band insulator.

Alas, this was not to be: the triangular $S=1/2$ antiferromagnet
exhibits N\'eel order. While a simple collinear
two-sublattice N\'eel structure is precluded by the non-bipartiteness
of the triangular lattice, the spin order distinguishes three
sublattices, on which the directions of the spin order parameter are
oriented at 120$^o$\cite{misguich-lhuillier}.

The question whether an RVB liquid could exist, as a matter of
principle, was left unanswered. Interest in this problem was greatly
intensified with the advent of high-temperature superconductivity and
the proposal that the superconducting phase effectively derived from
doping a parent RVB liquid \cite{AndersonScience}.

In the following, the RVB liquid continued to be elusive.  One
obstacle was that, when destabilising a bipartite N\'eel state, e.g.\
by adding frustrating interactions, the competing ground state would
typically tend to break some other local symmetry. A particular
prominent example is the valence bond solid, in which the order
parameter leaves the SU(2) invariance intact (as it involves bond
amplitudes $\langle {\bf S}_i\cdot{\bf S}_j\rangle$ rather than a
single spin, $\langle {\bf S}\rangle$), but nonetheless breaks
translational symmetry.

While no RVB liquid had thus become available, a proof of its
non-existence was not forthcoming either, and the search thus
continued. One route being followed was to study an increasing number
of Hamiltonians numerically in the hope of isolating the most
promising candidates \cite{misguich-lhuillier}. A second one was to
investigate analytically tractable models -- at the expense of quantum
chemical realisability -- and attempt a direct demonstration of the
existence of the RVB liquid. One possibility, for instance, consists
of enlarging the symmetry group to obtain a large-$N$ theory
\cite{rslargeN}, although the applicability to the small-$N$ regime remains 
unsettled.

In this presentation, we review how the second strategy can be carried
through entirely: we describe how, in a local SU(2) invariant $S=1/2$
Hamiltonian, one can demonstrate the existence of an RVB liquid. As a
byproduct, we can in fact show that a range of other interesting
valence bond dominated phases can similarly be shown to exist.

The two conceptually separate steps in this demonstration are (i) how
to transform the SU(2) Hilbert space into one of valence bonds by
using a local Hamiltonian and (ii) how to construct an effective
Hamiltonian in the reduced valence bond (`dimer') space that can be
demonstrated to lead to a liquid phase.

Step (ii) has as its starting point the seminal work by
Rokshar and Kivelson (RK) \cite{Rokhsar88}, who formulated their
quantum dimer model (QDM) in the hope of its exhibiting a liquid
phase. Whereas that hope was not fulfilled, it was later
shown \cite{MStrirvb} that the RK-QDM on the triangular lattice indeed
does exhibit such a liquid phase. A central ingredient in this
demonstration was the fact that at a particular point in parameter
space (the RK point), the RK-QDM is exactly soluble. It was thus
possible to demonstrate that all local correlators formulated in terms
of dimers are short-ranged. This was combined with numerical evidence
for the continuity of the physics at this point into an extended phase
(see in particular the recent work by the Lausanne group
\cite{lausanneGFMC}, also covered in this workshop).

The solution of step (i) builds on a thread of work initiated by
Klein \cite{Klein}, who wrote down a class of model Hamiltonians which
have nearest-neighbour valence bond coverings as their ground
states. This work was carried further by Chayes, Chayes and Kivelson
\cite{chayeschayeskivelson}, but, in the absence of the dimer liquid, 
activity in this direction waned.

\begin{figure}
  \includegraphics[height=.2\textheight]{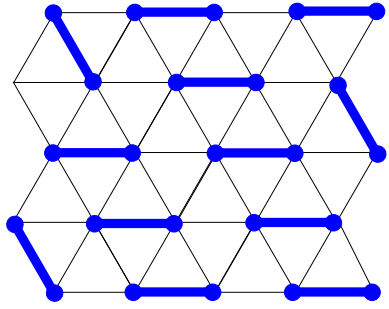}
  \caption{
A nearest-neighbour valence bond (dimer) covering for a triangular lattice.
}
  \label{fig:covering}
\end{figure}

In the following, we first concentrate on step (i). We
start with with a qualitative account of Klein models \cite{Klein},
and how they lead to valence bond ground states. In particular, we
argue that supplementary ground states can be excluded using a
decoration procedure, so that we are left with a low-energy sector of
degenerate nearest neighbour valence bond ground states, which we
refer to as dimer coverings. An example of such a covering is given in
Fig.~\ref{fig:covering}.

We then turn to the construction of effective quantum dimer
Hamiltonians within this subspace using the RK overlap expansion. We
construct operators which mimick the potential and kinetic terms of
the RK-QDM. We argue that the decoration procedure introduced above
can also be used to control the size of unwanted additional terms.
Readers interested in a detailed account of the technicalities are
referred to our original publication, Ref.~\cite{rmsklein}

This is followed by a brief account of the new phases attained by this
construction. These include the abovementioned SU(2) invariant RVB
liquid phase.

In closing the introduction, we would like to draw the reader's attention to a
complementary piece of work by Fujimoto \cite{fujimotoklein}, who constructs
Hamiltonians realising the RK-QDM for valence bond wavefunctions.

\section{Spin Hamiltonians as projectors}
The Heisenberg Hamiltonian
\begin{equation}
    H_{Heis}=
J\sum_{\langle i,j\rangle}\vec{s}_{i}\cdot\vec{s}_{j}
\end{equation}
can be considered to be a projector:
\bea
H_{Heis}=\sum_{\langle i,j\rangle} \hat{P}_{\{ij\}}^{S=1}
\nonumber
\eea
which exacts an energy $J$ if the pair of spins $\{i,j\}$ has total spin
$S=1$; a singlet $S=0$ costs no energy. 

The projectors $\hat{P}$ on different bonds do not commute, as one would
expect for a quantum model. Writing down a ground state for the full
Hamiltonian does therefore, in general, not reduce to minimising the
Hamiltonian for each bond separately. This is of course ultimately
what makes it so much more difficult to write down an exact ground
state for a quantum rather than a classical spin model.

The big insight of Klein was that there exists a half-way house. For
different operators to be simultaneously minimisable does not in fact
require that they commute. Rather, by a judicious choice of the
properties of the projectors, it can become possible to find a
subspace of the Hilbert space which is annihilated by {\em all} the
projectors, even though they do not commute. 

To achieve this, Klein proposed defining the projectors not for each
single bonds, as is the case for the usual Heisenberg model, but
instead for the neighbourhood $\mathcal{N}(i)$ of each site $i$. This
neighbourhood consists of a site and its $z-1$ nearest neighbours.
Again, the projector is constructed so that its exacts an energy cost
$J$ only when the spins in $\mathcal{N}(i)$ are in the maximally allowed
spin state, $S=z/2$. 

One can thus write the Klein Hamiltonian as
\begin{equation}
H_{K}=\sum_{i\in \Lambda}\hat{P}^{S=z/2}_{\mathcal{N}(i)}\ ,
\label{eq:basicklein}
\end{equation}
with $S$ given as
\begin{equation}
    \vec{S}_{\mathcal{N}(i)}=
\sum_{j\in\mathcal{N}(i)}\vec{s}_j
\end{equation}
This Hamiltonian corresponds not to a pair interaction as in the
Heisenberg case, but rather to a multispin interaction. The larger the
neighbourhood ${\mathcal{N}(i)}$, the more spins are involved in this
interaction. However, it always remains local.

For instance, for even $z$, one obtains
\begin{eqnarray}
    \hat{P}_{\mathcal{N}(i)}=
k_{i}\prod_{L=0}^{z_i/2-1}\Bigl[S_{\mathcal{N}(i)}^{2}-L(L+1)
\Bigr]\ .
\end{eqnarray}

It is now easy to see why this Hamiltonian is minimised by
nearest-neighbour valence bond coverings. Such coverings are defined
by complete pairings of the sites of the lattice such that each site
is paired to form a spin singlet with one of its nearest-neighbours.
This implies that the $z$ spins in ${\mathcal{N}(i)}$ can at most sum
up to a total spin $S=(z-2)/2$, and hence the corresponding projector
$\hat{P}_{\mathcal{N}(i)}$ is guaranteed to annihilate this state.

The simplest representative of this class of Hamiltonians is the
Majumdar-Ghosh model \cite{majghosh}, 
the one-dimensional chain in which spins
interact in neighbourhoods containing three spins, which gives it the
alternative appearance of a model with nearest and next-neaerest
neighbour interactions.

However, writing down such projectors in itself is of course not the
full story. After all, an even simpler Hamiltonian, $H\equiv0$, would
have had all valence bond coverings as ground states as well. If one
wants to obtain a quantum dimer model via the Klein route, it is
necessary to demonstrate that in fact there are no other ground states
for this Hamiltonian.

\begin{figure}
  \includegraphics[height=.2\textheight]{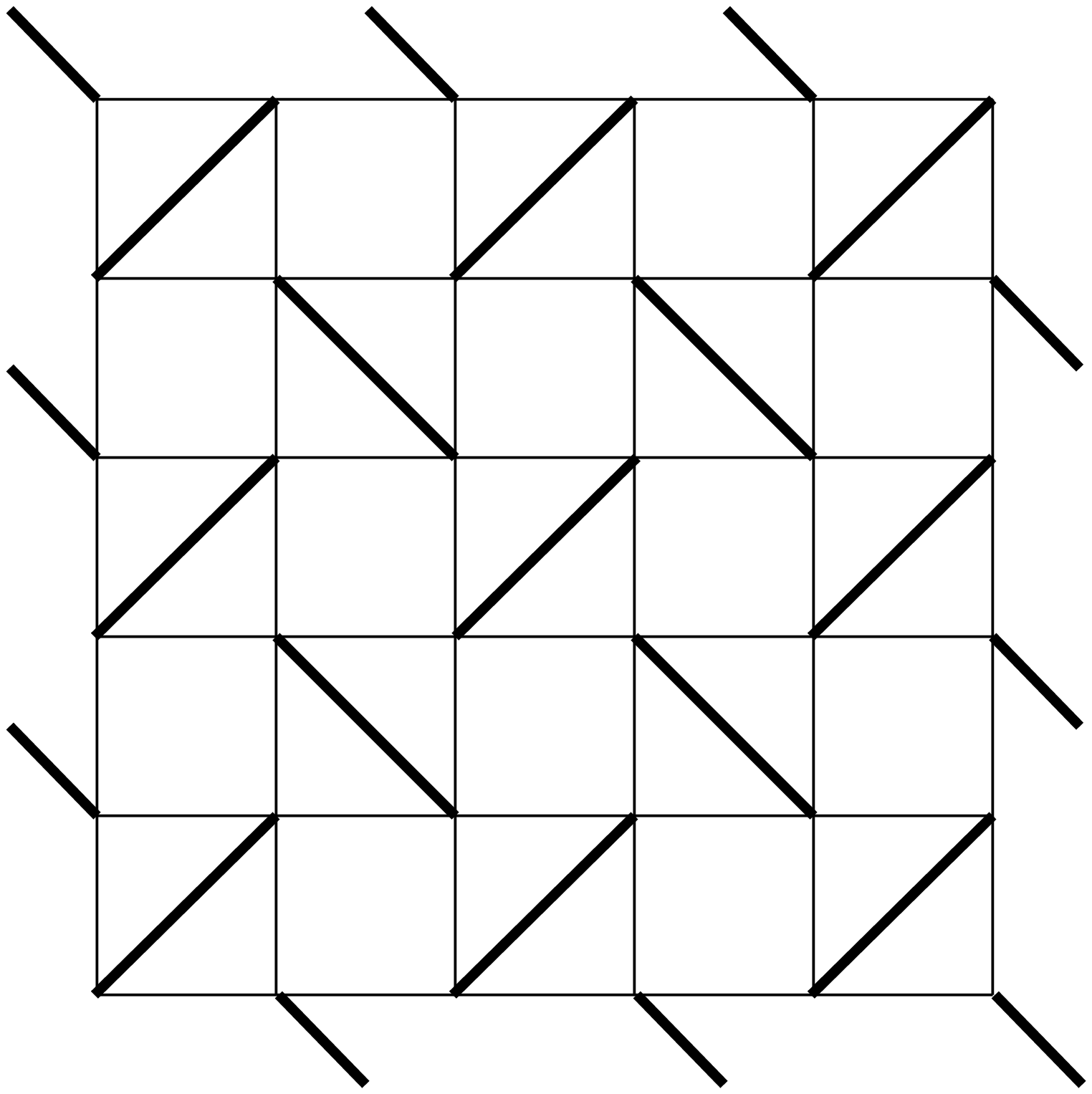}
\hfill
   \includegraphics[height=.2\textheight]{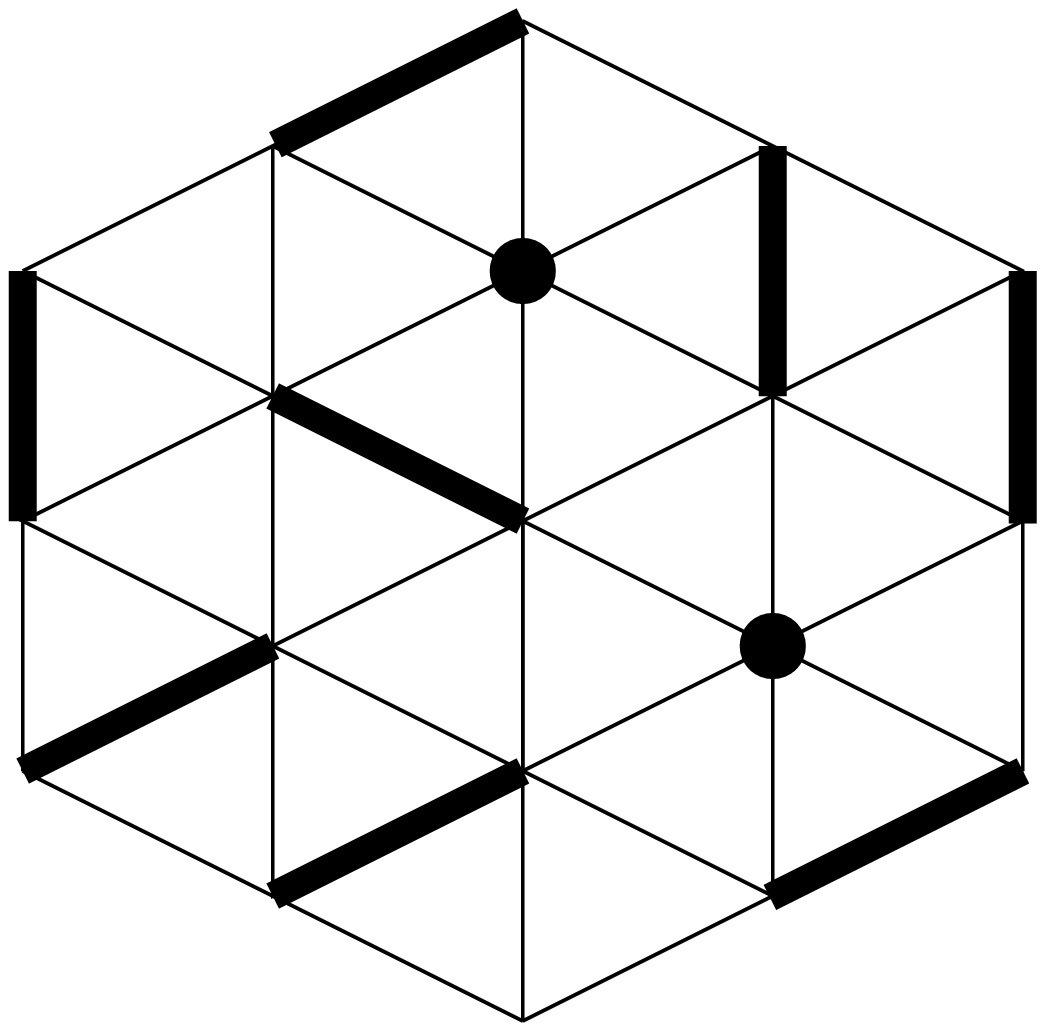}
  \caption{Ground states of the  Klein model which are not nearest neighbour
valence bond (dimer) coverings of the underlying lattice. These are removed
by the decoration procedure.}
  \label{fig:unwanted}
\end{figure}

In some cases, such a demonstration is possible, as for the case of
the honeycomb lattice \cite{chayeschayeskivelson}. In others, it is
possible to show explicitly that there are other ground states, as is
the case for the square or triangular lattices (see
Fig.~\ref{fig:unwanted}). For the Majumdar-Ghosh chain, Shastry and
Sutherland presented a calculation showing that excitations above the
dimerised ground states do remain gapped
\cite{ShasSutherMG}.

\section{Decoration}
\begin{figure}
  \includegraphics[height=.25\textheight]{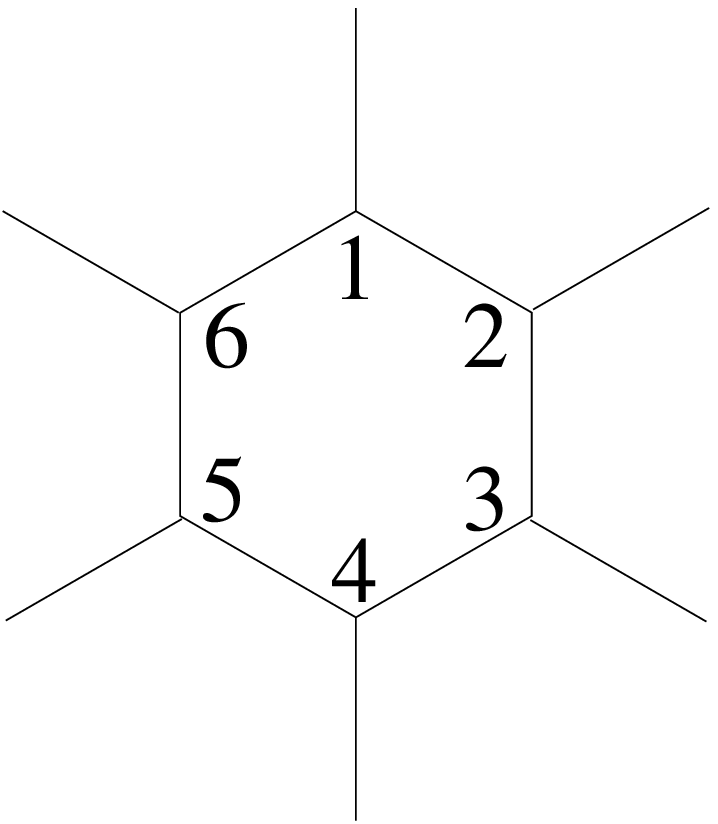}
\hfill
\psfrag{a1}{{a1}}
   \includegraphics[height=.25\textheight]{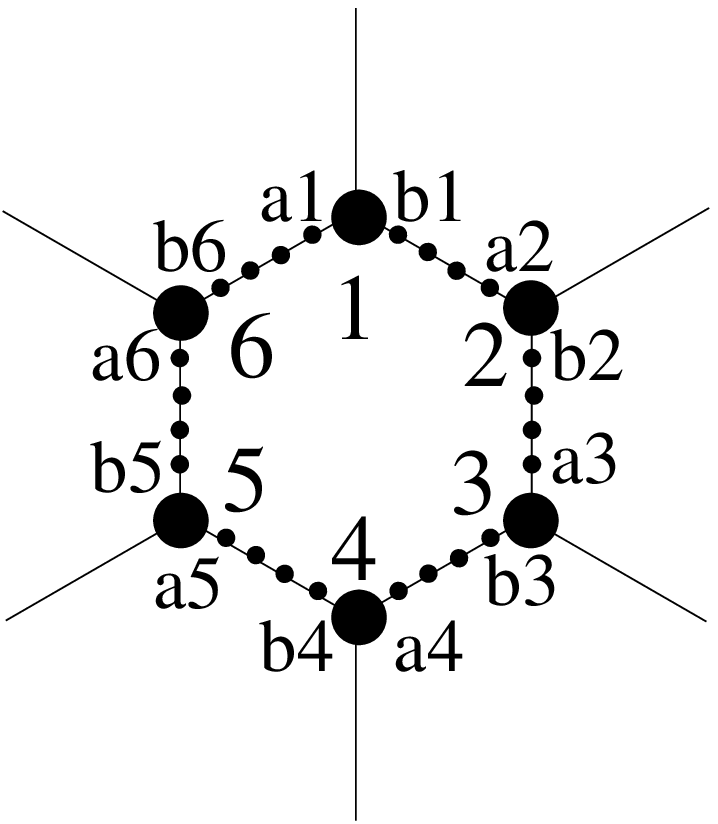} 
\caption{ Decoration
   procedure to remove unwanted ground states, and to obtain a control
   parameter for the RK overlap expansion. } 
\label{fig:decorate}
\end{figure}

This problem of the supernumerary ground states can be taken care of
by a decoration procedure (see Fig.~\ref{fig:decorate}). Details of
the necessary calculations are given in Ref.~\cite{rmsklein}. 

Basically, the decoration procedure places an {\em even} number, $N$,
of supplementary sites between each pair of sites of the original
lattice. The parity of the sites per unit cell is left unchanged by
this decoration procedure for a lattice of even coordination. In the
case of a lattice with odd coordination, one needs to choose $N$ to be
a multiple of 4 for this to be the case.

The Hamiltonian for the extended lattice is its Klein
Hamiltonian, with the possibility of varying the prefactor for the
projectors of a neighbourhood depending whether it contains a site of
the original lattice. 

For the supplementary sites, the projectors are of course of the
Majumdar-Ghosh type. Thus, thinking of the supplementary sites on a
given bond as a Majumdar-Ghosh chainlet (of finite length $N$), it is
clear that its interior will not harbour gapless excitation. 

Problems can arise close to the original sites, where several of these
chainlets meet, thus enabling any excitations to gain additional
kinetic energy as the local coordination is higher there.  Repeating
the Shastry-Sutherland calculation on the Majumdar-Ghosh model for the
present case, we find that it is possible to choose the
interaction parameters such that the gap is not destroyed. This
calculation is carried out explicitly for the case of the honeycomb
lattice in Ref.~\cite{rmsklein}.

Finally, one needs to note that the dimerisations of the decorated
lattice can be mapped bijectively onto those of the unprojected
one. This follows from the fact that there are only two possible
dimerisation patterns for the Majumdar-Ghosh chainlets. As we have
chosen $N$ to be even, one of these is compatible with dimers from the
original sites both pointing inwards into the chain, and the other
with no dimers pointing inwards. These correspond to the presence and
absence, respectively, of a dimer on the bond linking the two original
sites in the undecorated lattice. 

\section{Perturbation theory in the dimer manifold}
We now turn to the second step in the program -- how to get from a
valence bond Hilbert space to the desired dimer phases. In essence, we
are looking for terms in the Hamiltonian, subleading to the Klein
terms, which mimick the RK-QDM when projected onto the valence bond
ground state subspace.

The RK-QDM, in pictorial form for the square lattice (for which it was
written down originally \cite{Rokhsar88}), has the following form:
\begin{equation}
\scalebox{0.5}{\includegraphics{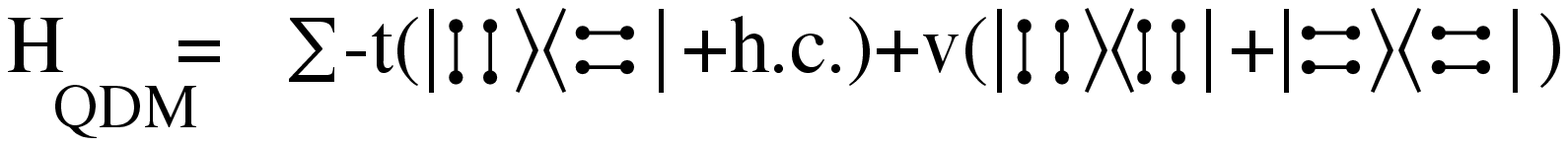}}\ .
\label{eq:RKQDM}
\end{equation}
It contains two terms, a kinetic and a potential one.  Firstly, the
resonance term  flips a pair of dimers around a plaquette; it
has a matrix element $-t$, where $t$ needs to be positive for the
quantum dimer model to be analytically tractable in a simple manner.
This can often, but by no means always, be arranged.

This term accounts for the resonance move to which the resonating
valence bond physics owes its name. The other is a potential term,
which exacts an energy cost $v$ for each plaquette that can
resonate. In practice, it important as (i) the RK point $v/t$ is often
soluble (or straightforwardly simulateable) and (ii) because it can be
used to counterbalance ordering tendencies, thus enhancing the
disordering effect of quantum fluctuations.

To give meaning to such a pictorial Hamiltonian, we first need to
specify precisely what we mean by the pictures of dimers, as the dimer
coverings are not orthogonal. Indeed, if we define coverings, $i$, of
the lattice with nearest-neighbour valence bonds, we can define an
overlap matrix $S$, with matrix elements
\bea
S_{ij}=\langle i|j\rangle\ ,
\eea
allowing us to define an orthonormal basis using the states $\alpha$
as follows:
\bea
|\alpha\rangle=\sum_{i}(S^{-1/2})_{\alpha,i}|i\rangle\ .
\label{eq:ortho}
\eea

The crucial observation is that, for our decorated lattice, the
overlap between two distinct dimer coverings is exponentially small in
$N$, the number of sites on the decorating Majumdar-Ghosh chainlets.
This follows from the observation that the overlap between two
coverings differing in a loop of $n$ dimers is $(1/2)^{n-1}$, and the
presence of the chainlets ensures that $n\propto N$.

We can thus identify a basis vector $\alpha$ with its leading
principal valence bond covering $i$, the admixture of other coverings
$j\neq i$ being exponentially suppressed.
In the orthonormalised basis, the matrix elements of the Hamiltonian
read
\bea
 H_{\alpha\beta}&=&(S^{-1/2}\delta H S^{-1/2})_{\alpha\beta}\\
    &=&\sum_{ij}(S^{-1/2})_{\alpha i}\langle i|\delta H|j\rangle
    (S^{-1/2})_{j\beta} \ .
\label{eq:overlapexp}
\eea
 
The calculation of each of the matrices appearing in this expression
involves the non-orthogonal dimer coverings, $i$, and can thus be
carried out relatively straightforwardly. 

For example, in the case of the honeycomb lattice, we use the
following perturbing Hamiltonian:

\bea
    \delta H &=& J \sum_{\langle ij \rangle} \vec{s}_{i}\cdot\vec{s}_{j} +
\nonumber\\
&&v
    \sum_{\hexagon}\Bigl(
    (\vec{s}_{1}\cdot\vec{s}_{b_{1}})(\vec{s}_{3}\cdot\vec{s}_{b_{3}})(\vec{s}_{5}\cdot\vec{s}_{b_{5}})\nonumber\\
&&    +
    (\vec{s}_{1}\cdot\vec{s}_{a_{1}})(\vec{s}_{3}\cdot\vec{s}_{a_{3}})(\vec{s}_{5}\cdot\vec{s}_{a_{5}})
    \Bigr)\ ,
\label{eq:perturb}
\eea
where the labeling of the sites is given in
Fig.~\ref{fig:decorate}. The understanding is that the energy
scales $v,J$ occuring here are much smaller than those of the Klein
model, so that it suffices to do degenerate perturbation theory.

Plugging the perturbing Hamiltonian (\ref{eq:perturb}) into the
overlap expansion (\ref{eq:overlapexp}) yields the following quantum
dimer Hamiltonian:
\bea
 H_{\alpha\beta}&=&-Jx^{6(N+1)}\hexagon_{\alpha\beta}+vn_{fl,\alpha}\delta_{\alpha\beta}\nonumber\\&+&O(vx^{6(N+1)}+Jx^{10(N+1)})\nonumber\\
    &=&-t\hexagon_{\alpha\beta}+vn_{fl,\alpha}\delta_{\alpha\beta}+O(vx^{6(N+1)}+tx^{4N}) \ .\label{eq:kleinqdm}
\eea
Here, $n_{fl}$ counts the number of plaquettes which can resonate in a
dimer configuration, and $\hexagon_{\alpha\beta}$ denotes a matrix
whose elements are non-zero if the the two dimer coverings differ only
by a single resonance move, and $x=1/\sqrt{2}$. These thus realise the
desired potential and kinetic terms, respectively.

As advertised before, the small parameter of our overlap expansion is
effectively $x^N$, which can be made arbitrarily small. We need to
point out that the interaction responsible for the potential term is
of range $N$ in units of nearest-neighbour distances of the decorated
lattice. Hence, for $N\rightarrow\infty$, our interaction would cease
to be local. However, the expectation is that the actual value of $N$
required for this scheme to work will not be all that large.

The benefit of the decoration scheme is therefore that it provides us
with a small control parameter which, unlike in the case of large-$N$
theories, is not related to an enlarged internal symmetry of spin
space: we are still dealing with the native SU(2) symmetry.

\section{Resulting valence bond phases}
Having established how to obtain a RK-QDM from an SU(2) invariant
Hamiltonian, we sketch in the following the type of phases which we
can realise in this way. The basic difference to bear in mind when
comparing to the work on pure RK-QDMs is that here we have additional
terms in the Hamiltonian, albeit of small size (controlled by the
decoration). Nonetheless, properties which require the fine-tuning of
some parameters, such as the existence of multicritical points, will
be sensitive to their presence, whereas a stable phase will
generically be robust.

\begin{figure}
\rotatebox{-90}{\includegraphics*[viewport = 150 00 516 842, height=.8\textheight]{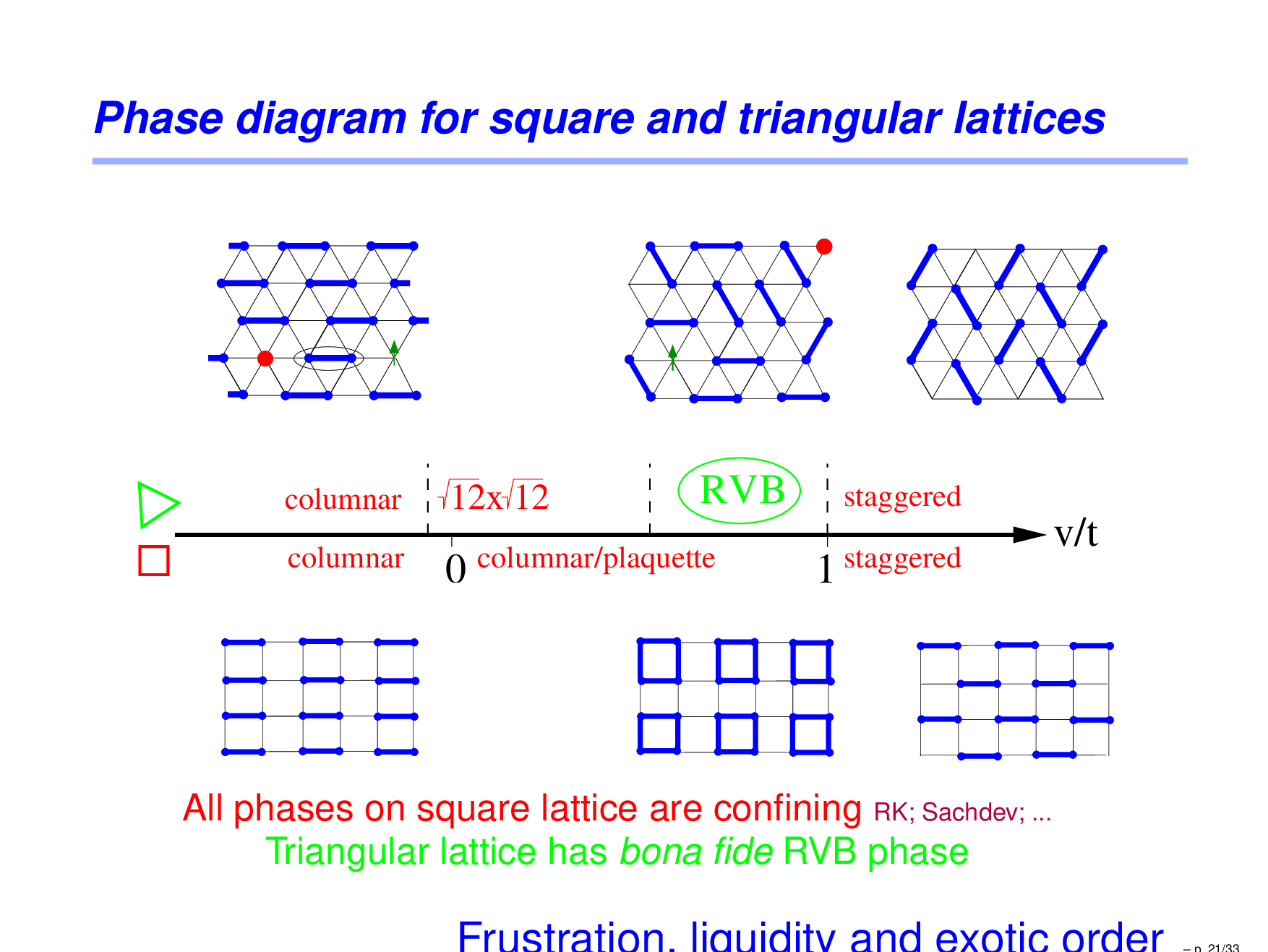}}
  \caption{ Schematic phase diagram of the RK-QDM on the square and
  triangular lattices.  The cartoons for the triangular lattice depict
  the presence of fractionalisation and deconfinement for the case of
  the RVB liquid phase, and its absence for the
  valence bond solid phases.  The RK-QDM model on the square lattice
  only has solid phases.  } \label{fig:qdmphase}
\end{figure}

\subsection{SU(2) invariant RVB liquids}

In Fig.~\ref{fig:qdmphase}, we show the phase diagram of the RK-QDM on
the triangular lattice. Its most salient feature is the presence of an
extended RVB liquid phase including the RK point and the vicinity to
its left. As this is a gapped phase, it will be stable to the
perturbations included in the Klein-derived quantum dimer Hamiltonian.
Our construction thus provides the point-of-principle demonstration
that SU(2) invariant RVB liquids do exist.

\subsubsection{Fractionalisation}
This RVB liquid corresponds to the deconfined phase of an effective
Ising gauge theory \cite{msf}. This is indicated in
Fig.~\ref{fig:qdmphase} in cartoon form. In the columnar phase,
separating two monomers creates a domain wall, the tension of which
leads to a diverging confining potential between the monomers as their
separation is increased.

By contrast, in the liquid phase, a monomer is oblivious to the
distance to its partner once their separation exceeds several
correlation lengths. The pair can thus be separated at a finite cost
in energy -- it is deconfined. 

\begin{figure}
 \includegraphics[height=.05\textheight]{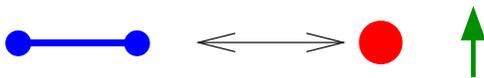}
\caption{ 
Removing an electron leaves behind a charged hole and an unpaired spin.
} 
\label{fig:monomerholon}
\end{figure}

This phenomenon is also known as spin-charge separation
\cite{AndersonScience}. This name is due to the observation that 
removing one electron leaves behind the partner with which it formed a
singlet bond (see Fig.~\ref{fig:monomerholon}). One is thus left with
a spinless charged hole and an uncharged spin. Separating these two at
finite cost in energy thus allows the independent existence of a
spin-0 charge-e object and a spin-1/2 charge-0 object. The phenomenon
of spin-charge separation is a particular instance of quantum number
fractionalisation.

\subsection{Cantor deconfinement}
The phase diagram for the square lattice RK-QDM is not the full story
as far as the quantum dimer model derived via the overlap expansion is
concerned. This is due to the small correction terms present in
Eq.~\ref{eq:kleinqdm}. Although exponentially small, they become
important close to the phase transition between plaquette and
staggered valence bond solid. This happens because the RK Hamiltonian
leads to a fix-point action for this phase transition which has a
symmetry which is higher than that dictated by the underlying
symmetries of the lattice. 

In other words, the phase transition in the RK-QDM corresponds to a
fine-tuned multicritical point, which fine tuning is undone by the
correction terms \cite{cantor}. We note in passing that this
multicritical point is very interesting in its own right. It is an
example of a critical point exhibiting deconfinement, whereas the
phases it separates are both confined \cite{MStrirvb}. The phenomenon
of such `deconfined quantum criticality' in a more general setting has
received a great deal of attention recently \cite{dqcp}. For an 
interesting example of a deconfined critical point in a microscopic
model, see Ref.~\cite{alexeidqcp}.

\begin{figure}
\psfrag{C}{\small{columnar}}
\psfrag{P}{\small{plaquette}}
\psfrag{I}{\small{(in)commensurate}}
\psfrag{T}{\small{tilt}}
\psfrag{m}{\small{max}}
\psfrag{R}{\small{RK point}}
\psfrag{0}{\small $0$}
\psfrag{inf}{\small $-\infty$}
\psfrag{1}{\small $1$}
\psfrag{pinf}{\small $\infty$}
\psfrag{s}{\small{staggered}}
\psfrag{v}{\small $v/t$}
\includegraphics[width=0.95\textwidth]{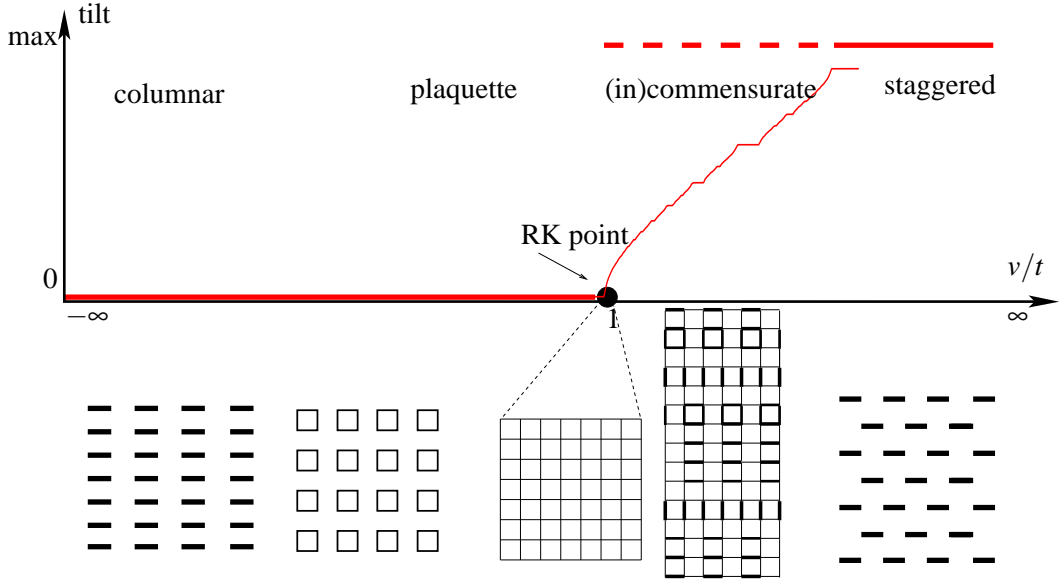}
\caption{The phase diagram of the generalised RK-QDM on the
square lattice. A detailed analysis is given in Ref.~\cite{cantor}.
The vertical axis, 'tilt', is zero in the columnar and plaquette
phases, and maximal (i.e.\ limited by lattice effects) in the
staggered phase. Its jump in the RK model, where the RK point is
fine-tuned multicritical, is replaced by an ascending devil's
staircase in the Klein-derived quantum dimer model.}
\label{fig:cantorphase}
\end{figure}

The phase diagram resulting instead is shown in
Fig.~\ref{fig:cantorphase}. There, the abrupt change from plaquette to
staggered valence bond solid is replaced by a continuous growth of an
appropriately defined order parameter, the `tilt' \cite{cantor}. Due
to an interplay between a locking potential due to the underlying
lattice and the tilt favoured in its absence (which tilt may be
incommensurate), the growth is not smooth, but occurs in the form of a
devil's staircase. Provided that no first-order phase transition
intervenes, the region close to the RK point is effectively
deconfined, a phenomenon we have termed Cantor deconfinement -- for
the small print, see Ref.~\cite{cantor}.

\subsection{Artificial electromagnetism}
The long-wavelength description of the QDM on the square lattice is
that of a U(1) gauge theory in $d=2+1$. The presence of
confinement everywhere, well known from high energy physics, is a
consequence of this structure. This makes the existence of the Cantor
deconfined region somewhat remarkable. However, in essence this is
possible because it occurs in a sector of the gauge theory not covered
by the conventional wisdom.

In three dimensions, things are considerably simpler as the existence
of a deconfined phase in the corresponding U(1) gauge theory in
$d=3+1$ is not precluded on general grounds. In fact, the
three-dimensional dimer model, obtained along the lines discussed
above, is expected to exhibit a Coulomb phase in which quasiparticles
fractionalise. Perhaps just as strikingly, this phase supports
transverse collective excitations which are completely analogous to
photons in conventional electromagnetism
but which represent an emergent excitation: they represent a
collective excitation of the SU(2) spins-1/2 on the simple cubic
lattice. Endowed with the Klein Hamiltonian, this system can act as
ether for an artificial electromagnetism \cite{artificiallight}.

\begin{theacknowledgments}
R. M. would like to thank the organisers of the workshop in Peyresq
for inviting him to participate. He is grateful to the Aspen Center
for Physics, where parts of this work were undertaken.  This work was
in part supported by the Minist\`ere de la Recherche et des Nouvelles
Technologies with an ACI grant and by the NSF (grant DMR-0213706).
\end{theacknowledgments}


 \end{document}